\documentclass[prd,aps,showpacs,nofootinbib,nobibnotes,twocolumn,graphicx,psfrag]{revtex4}
\usepackage{epsfig}

\newcommand{\dd}{{\rm d}}

\begin{document}
\title{The distance duality relation from X-ray and SZ observations of clusters}

 \author{Jean-Philippe Uzan}
 \email{uzan@iap.fr,uzan@th.u-psud.fr}
 \affiliation{Institut d'Astrophysique de Paris, GR$\varepsilon$CO,
             98bis boulevard Arago, 75014 Paris, France \\
             Laboratoire de Physique Th\'eorique, CNRS-UMR 8627,
             Universit\'e Paris Sud, B\^atiment 210, F--91405 Orsay
             c\'edex, France}
 \author{Nabila Aghanim}
 \email{aghanim@ias.u-psud.fr}
 \affiliation{Institut d'Astrophysique Spatiale,
             Universit\'e Paris Sud, B\^atiment 121, F--91405 Orsay
             c\'edex, France\\
             Division of Theoretical Astronomy,
             National Astronomical Observatory of Japan,
             Osawa 2-21-1, Mitaka, Tokyo 181-8588, Japan}
 \author{Yannick Mellier}
 \email{mellier@iap.fr}
 \affiliation{Institut d'Astrophysique de Paris,
             98bis boulevard Arago, 75014 Paris, France\\
             Observatoire de Paris, LERMA,
             61 av. de l'observatoire, 75014 Paris, France.}

\date{May 25, 2004}
\begin{abstract}
X-ray and Sunyaev-Zel'dovich data of clusters of galaxies enable
to construct a test of the distance duality relation between the
angular and luminosity distances. We argue that such a test on
large cluster samples may be of importance while trying to
distinguish between various models of dark energy. The analysis of
a data set of 18 clusters shows no significant violation of this
relation. The origin and amplitude of systematic effects and the
possibility to increase the precision of this method are
discussed.
\end{abstract}
\maketitle

\section{Introduction}

Most cosmological observations provide compelling evidences that
our universe is undergoing a late time acceleration
phase~\cite{dark_revue}. However, there are still several debates
about the physical interpretation of these observations. While it
seems clear that the Friedmann equations for a universe only
composed of normal matter (i.e. radiation and dust) even including
dark matter cannot explain the current data, there are different
ways of facing this fact. Either one can conclude that the
interpretation of the cosmological data are not correct (i.e. we
do not accept the evidence for the acceleration of the universe,
see Ref.~\cite{dark_revue} for a recent critical review and e.g.
Ref.~\cite{pascc}) or one tries to introduce new degrees of
freedom in the cosmological model. In this latter case, these
extra degrees of freedom, often referred to as {\it dark energy},
can be introduced as a new kind of matter or as a new property of
gravity.

In the first approach one assumes that gravitation is described by
general relativity while introducing new forms of gravitating
components, beyond the standard model of particle physics, which
must have some effective negative pressure to explain the
acceleration of the universe. Various candidates such as a
cosmological constant, quintessence~\cite{dark_revue,quint} with
many potentials, K-essence~\cite{kes} etc... have been proposed.
But, one is still left with the cosmological constant
problem~\cite{ccp} (why is the density of vacuum energy expected
from particle physics so small?) as well as the time coincidence
problem (why does the dark energy starts dominating today?)
unsolved. From a cosmological point of view, these models are
characterized by their equation of state which can be
reconstructed from the function $E(a)=H^2(a)/H_0^2$ where $H_0$ is
the Hubble constant at present and $a$ the scale factor, either
using the observation of background quantities or the growth of
cosmic structures~\cite{bb}.

\begin{figure*}[t]
 \centerline{\epsfig{figure=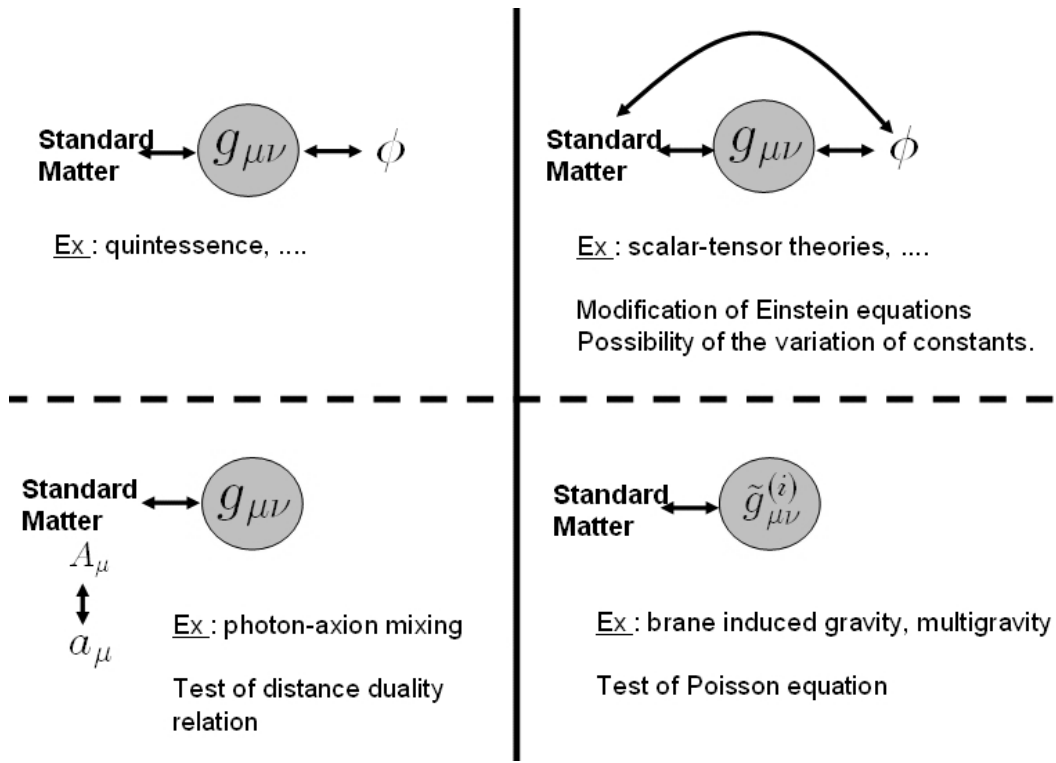,width=14cm}}
 \caption{Summary of the different classes of models and of the specific tests
 that can help distinguish between them (besides the equation of state
 and the growth of cosmic structures). The classes differ according
 to the kind of new fields and to the way they couple to the metric
 $g_{\mu\nu}$ and to the standard matter fields. Upper-left class consists of
 models in which a new kind of gravitating matter is introduced,
 e.g. quintessence. In the upper-right
 class, a light field induces a long-range force so that gravity is
 not described by a spin-2 graviton only. This is the case of
 scalar-tensor theories of gravity. In this class, Einstein equations
 are modified and there may be a variation of the fundamental
 constants. The lower-right class corresponds to models in which there
 may exist massive gravitons, such as in some class of braneworld
 scenarios. These models predict a modification of the Poisson
 equation on large scales. In the last class (lower-left), the
 distance duality relation may be violated.}
 \label{fig0}
\end{figure*}

The other route is to allow for modification of gravity. This
means that the only long range force that cannot be screened is
assumed to be not described by general relativity. Once such a
possibility is considered, many classes of models exist~(see e.g.
Ref.~\cite{will}). For instance, a light scalar field can couple
to matter leading to models of extended quintessence~\cite{uzan99}
and more generally to scalar-tensor types of theories. Such
theories have some difficulties to explain the current
cosmological observations~\cite{pef} without a quintessence-like
potential for the scalar field or a cosmological constant. This
scalar field may also be at the origin of some variation of the
fundamental constants, depending on its couplings, and violation
of the universality of free fall (see Ref.~\cite{uzan02} for a
review). Other possibilities include braneworld models in which
the standard model fields are localized on a 3-dimensional brane
embedded in a higher dimensional spacetime.  Among braneworld
models, a subclass of models have the property of allowing for
deviation from 4-dimensional Einstein gravity on large scales.
This is for example the case of some multi-brane
models~\cite{multi_brane}, multigravity~\cite{multigravity}, brane
induced gravity~\cite{induced_grav} or simulated
gravity~\cite{simulated_grav}. In such models, gravity is not
mediated only by massless gravitons, one therefore expects to have
deviations from Newton inverse square law on large scales. Testing
the Poisson equation on large scales may be a way to distinguish
between these alternatives~\cite{ub,uzan03}.

The different types of models are summarized schematically on
Fig.~\ref{fig0}. A diagnostic of the cause of the acceleration of
the universe will require to make many tests. In particular, the
reconstruction of the function $E(a)$ (or equivalently measuring
the effective equation of state of the dark energy) will not be
sufficient to distinguish between many models. It is thus
important to simultaneously check for the Poisson equation, the
growth of structure and the variation of the constants.

\begin{figure}
 \centerline{\epsfig{figure=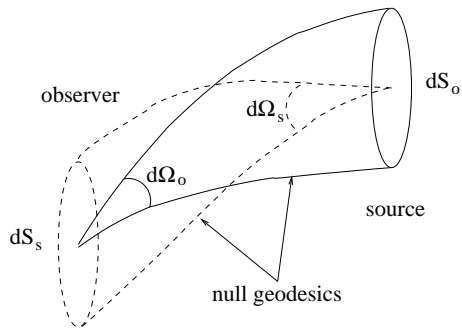,width=6cm}}
 \caption{A bundle of null geodesics diverging from the source (dash)
 subtending a solid angle $\dd\Omega_{\rm s}$ has a cross section $\dd S_{\rm s}$ at the observer
 while a bundle converging at the observer (plain) subtending a solid angle
 $\dd\Omega_{\rm o}$ has a cross section $\dd S_{\rm o}$ at the source.}
 \label{fig0bis}
\end{figure}

Among these tests it has recently been pointed out in
Ref.~\cite{basset_kunz} that the reciprocity relation and the
distance duality relation that derives from it have also to be
checked. The {\it reciprocity relation} is a relation between the
source angular distance, $r_{\rm s}$, and the observer area
distance, $r_{\rm o}$. The former is defined by considering a
bundle of null geodesics diverging from the source and which
subtends a solid angle $\dd\Omega_{\rm s}$ (see
Fig.~\ref{fig0bis}). This bundle has a cross section $\dd S_{\rm
s}$ and the source angular distance is defined by the relation
\begin{equation}
\dd S_{\rm s}=r_{\rm s}^2\dd\Omega_{\rm s}.
\end{equation}
The observer area distance $r_{\rm o}$ is defined analogously by
considering a null geodesic bundle converging at the observer by
\begin{equation}
\dd S_{\rm o}=r_{\rm o}^2\dd\Omega_{\rm o}.
\end{equation}
It can be shown that if photons travel along null geodesics and
the geodesic deviation equation holds then these two distances are
related by the reciprocity relation (see Ref.~\cite{ellis71} for a
derivation)
\begin{equation}
 r_{\rm s}^2 = r_{\rm o}^2 (1+z)^2,
\end{equation}
regardless of the metric and  matter content of the spacetime.
Unfortunately, the solid angle $\dd\Omega_{\rm s}$ cannot be
measured so that $r_{\rm s}$ is not an observable quantity. But,
it can be shown that, if the number of photons is conserved, the
source angular distance is related to the luminosity distance,
$D_L$, by the relation~\cite{ellis71}
\begin{equation}
 D_L=r_{\rm s}(1+z).
\end{equation}
It follows that there exist a {\it distance duality relation}
\begin{equation}\label{reci}
 {D_L}={D_A}(1+z)^2
\end{equation}
that holds between the angular distance $D_A$, the luminosity
distance $D_L$ and the redshift $z$. This relation can be checked
observationally .

While the reciprocity relation holds as soon as photons follow
null geodesic and that the geodesic deviation equation is valid,
the distance duality relation will hold if the reciprocity
relation is valid and the number of photon is conserved. In fact,
one can show that in a metric theory of gravitation, if Maxwell
equations are valid, then both the reciprocity relation and the
area law are satisfied and so is the distance duality relation
(see Ref.~\cite{ellis71}).

There are many possibilities for one of these conditions to be
violated. For instance the non-conservation of the number of
photons can arise from absorption by dust, but more exotic models
involving photon-axion oscillation in an external magnetic
field~\cite{axion} can also be a source of violation~\cite{kb2}.
Note also that in principle both the reciprocity and distance
duality relations hold for infinitesimal light bundles so that
gravitational lensing may be a source of violation for macroscopic
bodies. More drastic violations would arise from theories in which
gravity is not described by a metric theory and in which photons
do not follow null geodesic.

In this paper, we propose and explore a potential new  test of the
distance duality relation based on  Sunyaev-Zel'dovich~\cite{sz}
(SZ) and X-ray measurements of clusters of galaxies. In
Section~\ref{sec_2}, we first show that when the relation
(\ref{reci}) does not hold, cluster data do not give a measurement
of the angular distance but of $D_A/\eta^2$ where
\begin{equation}\label{def_eta}
 \eta(z) \equiv \frac{D_A}{D_L}(1+z)^2.
\end{equation}
In Section~\ref{sec_3}, we use existing cluster data to search for
any hint that $\eta=1$ may be excluded and then we discuss in
Section~\ref{sec_4} the possibility to improve the accuracy of the
test.

\section{Reciprocity relation from galaxy cluster
observations}\label{sec_2}

Galaxy clusters are known as the largest gravitationally bound
systems in the universe. They contain large quantities of hot and
ionized gas which temperatures are typically $10^{7-8}$~K. The
spectral properties of intra-cluster gas show that it radiates
through bremsstrahlung in the X-ray domain. Therefore, this gas
can modify the  Cosmic Microwave Background (CMB) spectral energy
distribution through inverse Compton interaction of photons with
free electrons. This is the so-called SZ effect. It induces a
decrement in the CMB brightness at low frequencies and an
increment at high frequencies.

The possibility of using the SZ effect together with  X-ray
emission of galaxy clusters to measure angular distances was
suggested soon after the SZ effect was pointed out (see for
example Ref.~\cite{sw}). Used jointly, they provide an independent
method to determine distance scales and thus to measure the value
of the Hubble constant (e.g. Ref.~\cite{bha,inagashi} for
details).

In brief, the method is based on the fact that the CMB temperature (i.e.
brightness) decrement due to the SZ effect is given by
\begin{equation}\label{eqne}
\Delta T_{\rm SZ} \sim L\overline{n_eT_e}
\end{equation}
where the bar refers to an average over the line of sight and $L$ is
the typical size of the line of sight in the cluster. $T_e$ is the
electron temperature and $n_e$ the electron density. Besides, the
total X-ray surface brightness is given by
\begin{equation}
 S_X \sim \frac{V}{4\pi D_L^2}\overline{n_en_pT_e^{1/2}}
\end{equation}
where the volume $V$ of the cluster is given in terms of its
angular diameter by $V=D_A^2\theta^2L$. It follows that
\begin{equation}
 S_X \sim \frac{\theta^2}{4\pi}\frac{D_A^2}{D_L^2}L
 \overline{n_en_pT_e^{1/2}}.
\end{equation}
The usual approach~\cite{sw}, is to assume the distance duality
relation ($\eta=1$) so that forming the ratio $\Delta T_{\rm
SZ}^2/S_X$ eliminates $n_e$.  Then, using a measurement of the
angular diameter of the cluster and Eq.~(\ref{thetac}) one gets an
estimate of the angular diameter distance and thus the Hubble
constant.  As a first conclusion, we point out that this method
determines the angular distance only if the distance duality
relation is valid. Therefore one needs to be careful when using
such data to test the distance duality relation.

To make this point more precise, let us come back to the details of
the method assuming the classical $\beta$-model for the galaxy cluster
\cite{king}, that is assuming that the electron density of the hot
intra-cluster gas has a profile of the form
\begin{equation}\label{eqprof}
 n_e(r) =
 n_0\left[1+\left(\frac{r}{r_c}\right)^2\right]^{-3\beta/2}.
\end{equation}
for $0<r<R_{\rm cluster}$ and 0 otherwise, $R_{\rm cluster}$ being
the maximum extension of the cluster. The temperature decrement
due to the SZ effect in the Rayleigh-Jeans part of the spectrum is
given by
\begin{equation}
 \Delta T_{\rm SZ}(\theta) =
 -2\frac{kT_0}{m_ec^2}\sigma_T\int_{-\ell_{\rm max}}^{\ell_{\rm max}}
 n_e \dd\ell
\end{equation}
where we have assumed that the temperature of the hot gas, $T_e$,
is independent of $r$ [$T_0\equiv T_e(r=0)$]. $2\ell_{\rm max}$ is
the length of the path along the line of sight inside the halo of
the cluster and $\theta$ is the angular radial position
 projected on the celestial sphere from
 the cluster center. The X-ray emission is due to thermal
bremsstrahlung and the surface brightness in a beam of angular
diameter $\delta\theta$ takes the form
\begin{equation}
 S_X(\theta) =
 \frac{\delta\theta^2}{4\pi}\frac{D_A^2}{D_L^2}\int_{-\ell_{\rm
 max}}^{\ell_{\rm max}} \frac{\dd L_X}{\dd V}\dd\ell
\end{equation}
where the emissivity in the frequency band $[\nu_1,\nu_2]$ is
given by
\begin{equation}
 \frac{\dd L_X}{\dd V} = \alpha(T_e,\nu_1,\nu_2,z)n_e^2.
\end{equation}
$\alpha(T_e,\nu_1,\nu_2,z)$ is a function that depends on the
properties of the free-free emission for ions, on the mass
fraction of hydrogen and on the gas temperature (see e.g.
Refs~\cite{bha,inagashi} for its expression). Introducing the
angle $\theta_c$ by
\begin{equation}\label{thetac}
 \theta_c=r_c/D_A,
\end{equation}
where $r_c$ is the cluster core radius, and using the profile
(\ref{eqprof}) we obtain in the limit $R_{\rm
cluster}\rightarrow\infty$
\begin{eqnarray}\label{eq13}
 \Delta T_{SZ}(\theta) &=& -2\frac{kT_0}{m_ec^2}\sigma_Tn_0r_c
 B\left(\frac{3\beta-1}{2},\frac{1}{2}\right)\nonumber\\
 &&\qquad\quad\left[1+\left(\frac{\theta}{\theta_c}\right)^2\right]^{(1-3\beta)/2}
\end{eqnarray}
and
\begin{eqnarray}\label{eq14}
 S_X(\theta) &=&
 \frac{\delta\theta^2}{4\pi}\frac{D_A^2}{D_L^2}\alpha n_0^2r_c
 B\left(\frac{6\beta-1}{2},\frac{1}{2}\right)\nonumber\\
 &&\qquad\quad
 \left[1+\left(\frac{\theta}{\theta_c}\right)^2\right]^{(1-6\beta)/2},
\end{eqnarray}
where $B$ is the Euler beta function. Using the definition of
$\eta$ from Eq.~(\ref{def_eta}), this latter expression rewrites
as
\begin{eqnarray}\label{eq14bis}
 S_X(\theta) &=&
 \frac{\delta\theta^2}{4\pi}\frac{\eta^2(z)}{(1+z)^4}\alpha n_0^2r_c
 B\left(\frac{6\beta-1}{2},\frac{1}{2}\right)\nonumber\\
 &&\qquad\quad
 \left[1+\left(\frac{\theta}{\theta_c}\right)^2\right]^{(1-6\beta)/2}.
\end{eqnarray}
As expected, $\Delta T_{\rm SZ}^2/S_X$ eliminates $n_e$ and gives
a measurement of the core radius $r_c$ from which we can deduce
the angular diameter distance through Eq.~(\ref{thetac}). When
$\eta\not=1$, what is thus extracted from the data is an estimate
of $\widetilde r_c=r_c/\eta^2$.

It follows from this analysis that, if we do not assume the
distance duality relation to hold, what is in fact determined is
$D_A^{\rm data}(z)=\widetilde r_c/\theta_c$ which differs from the
angular distance. We thus have access to
\begin{equation}
 D_A^{\rm data}(z)=D_A(z)/\eta^2(z)
\end{equation}
which reduces to the angular diameter distance only when the
distance duality relation holds.

\section{Method and data analysis}\label{sec_3}

Our method is straightforward once we have made the previous
remark. Using a data set of angular distances determined from the
combination of X-ray and SZ measurements, we have access to
$\lbrace z,D_A(z)/\eta^2(z)\rbrace$. To get $\eta$ one needs to
know the angular diameter distance. One possibility, and probably
the most robust, is to estimate it from its theoretical expression
in a Friedmann-Lema\^\i{t}re universe
\begin{equation}
 D_A^{\rm Th}(z) = f_K\left[\int_{1/(1+z)}^1\frac{\dd x}{x^2E(x)}\right]
\end{equation}
where $x=1/(1+z)$ and $f_K$ is defined by
\begin{equation}
 f_K(u)=\left(\frac{\sin\sqrt{K}u}{\sqrt{K}},u,\frac{\sinh\sqrt{-K}u}{\sqrt{-K}}\right)
\end{equation}
respectively for $K=H_0^2\left(1-\Omega_{\rm mat}^0 -
\Omega_\Lambda^0\right)/c^2$ positive, null and negative. The
function $E^2(x)=H(x)/H_0$ is explicitly given, for a
$\Lambda$-CDM model by
\begin{equation}
E^2(x) = \Omega_{\rm mat}^0 x^{-3} + \Omega_\Lambda^0 +
\left(1-\Omega_{\rm mat}^0 - \Omega_\Lambda^0\right) x^{-2},
\end{equation}
where $\Omega_{\rm mat}^0$ and $\Omega_\Lambda^0$ are respectively
the present matter and cosmological constant density parameters.

We estimate $\eta(z)$ as
\begin{equation}\label{calc_eta}
 \eta(z)=\sqrt{D_A^{\rm Th}/D_A^{\rm data}}.
\end{equation}
The error bars on this quantity will be estimated by a combination
of the data error bars and of the 1$\sigma$ error bars on the
cosmological parameters as obtained from Ref.~\cite{wmap}
\begin{equation}
 \Omega_{\rm mat}^0=0.29\pm0.07,\quad
 \Omega_\Lambda^0=0.73\pm0.05,\quad
 h=0.73\pm0.04.
\end{equation}
Note that in the case of a detection of $\eta\not=1$, the
interpretation of the signal is not trivial since a varying
equation of state may be undistinguishable from a violation of the
distance duality relation. In such a case going back to SNIa data
to get $D_L$ and of X-ray data to get $D_A/\eta^2$ may help to
break this degeneracy.

\begin{figure}
 \centerline{\epsfig{figure=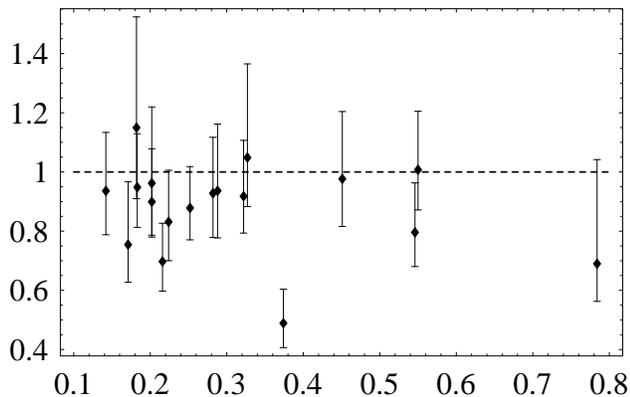,width=8.5cm}}
 \caption{$\eta$ as a function of the redshift for the 18 clusters
 of the Reese {\em et al.}~\cite{reese} catalog. The error bars
 include the observational error bars as determined by Reese {\em et
 al.} and the uncertainties in the cosmological parameters.}
 \label{fig1}
\end{figure}

We use the catalog by Reese {\em et al.}~\cite{reese} that
contains 18 galaxy clusters, with redshifts ranging from 0.142 to
0.784, all observed in X-ray and SZ (see Table~\ref{table1}).
Combining these data as discussed in Sect. \ref{sec_2} together
with the theoretical estimate of the angular distance, we get a
measurement of $\eta(z)$ for each cluster, using
Eq.~(\ref{calc_eta}). The result is summarized on
Table~\ref{table1} and is depicted on Figure~\ref{fig1}.

\begin{table}
\begin{tabular}{lccc}
 \hline\hline
 cluster & $\,$redshift$\,$ & $D_A^{\rm data}$ (Mpc) & $\eta$ \\
 \hline
 MS $1137.5+6625$ & 0.784 & $3179 ^{+1103}_{-1640}$ & $0.689_{-0.127}^{+0.352}$\\
 MS $0451.6-0305$ & 0.550 & $1278 ^{+265 }_{-299 }$   & $1.001_{-0.136}^{+0.198}$\\
 Cl $0016+16$ & 0.546 & $2041 ^{+484 }_{-514 }$ & $0.796_{-0.116}^{+0.167} $ \\
 RX J$1347.5-1145$ & 0.451 & $1221 ^{+368 }_{-343 }$ & $0.977_{-0.161}^{+0.227}$ \\
 Abell 370 & 0.374 & $4352 ^{+1388}_{-1245}$ & $ 0.489_{-0.083}^{+0.115}$ \\
 MS $1358.4+6245$ & 0.327& $  866 ^{+248 }_{-310 }$& $1.049_{-0.166}^{+0.316}$  \\
 Abell 1995 & 0.322 & $1119 ^{+247 }_{-282 }$ & $0.918_{-0.124}^{+0.189}$ \\
 Abell 611 & 0.288 & $  995 ^{+325 }_{-293 }$ & $0.936_{-0.159}^{+0.225}$ \\
 Abell 697 & 0.282 & $  998 ^{+298 }_{-250 }$ & $ 0.928_{-0.149}^{+0.189}$ \\
 Abell 1835 & 0.252 & $1027 ^{+194 }_{-198 }$ & $ 0.878_{-0.108}^{+0.140}$ \\
 Abell 2261 & 0.224 & $1049 ^{+306 }_{-272 }$ & $ 0.831_{-0.131}^{+0.174}$ \\
 Abell 773    & 0.216 & $1450 ^{+361 }_{-332 }$& $ 0.697_{-0.010}^{+0.129}$ \\
 Abell 2163    & 0.202 & $  828^{+181 }_{-205 }$ & $ 0.899_{-0.119}^{+0.179}$ \\
 Abell 520    & 0.202 & $  723 ^{+270 }_{-236 }$ & $ 0.962_{-0.176}^{+0.258}$ \\
 Abell 1689    & 0.183 & $  688 ^{+172 }_{-163 }$  & $ 0.948_{-0.136}^{+0.181}$ \\
 Abell 665    & 0.182 & $  466 ^{+217 }_{-179 }$ & $1.149_{-0.240}^{+0.374} $ \\
 Abell 2218  & 0.171 & $1029 ^{+339 }_{-352 }$ & $0.754_{-0.127}^{+0.213} $ \\
 Abell 1413 & 0.142 & $  573 ^{+171 }_{-151 }$ &  $0.936_{-0.148}^{+0.198} $ \\
\hline\hline
\end{tabular}
 \caption{The 18 clusters of the Reese catalog with their redshift used in our analysis.
 $D_A^{\rm data}$ refers to the angular distance determined in Ref.~\cite{reese} assuming
 that the distance duality relation holds. It leads, once this hypothesis is
 relaxed, to a measurement of $\eta$ with 1$\sigma$ error bars.}
 \label{table1}
\end{table}

The question is then whether this data set is compatible with
$\eta=1$ or not. As can be seen from the original data (see
Ref.~\cite{reese}), the error bars on $D_A^{\rm data}$ are not
symmetric. To derive the distribution of $\eta$ we proceed in the
following way. We first assume that the data points are
independent so that the likelyhood $L=P(\eta_1^{\rm
data}\ldots\eta_n^{\rm data}\vert\eta)$ can be factorized as
$L=\prod_i P_i(\eta_i^{\rm data}\vert\eta)$. We then introduce $S$
defined as
\begin{equation}
 S = -2\ln L.
\end{equation}
If the probabilities $P_i$ are Gaussian, $S$ reduces to the
standard $\chi^2$. In one dimension, a variation $\Delta S=1$
around the minimum of $S$ will give the $1\sigma$ error bar. To
proceed, we need to know the probabilities $P_i$. Without any
further information, we assume that they follow a Gaussian
distribution that is that ${\eta_i}^{+\delta_i^+}_{-\delta_i^-}$
corresponds to a probability distribution function of the form
\begin{equation}
 P(x)=\sqrt{2\over\pi}\frac{1}{(\delta_i^+ + \delta_i^-)}\left\lbrace
 \begin{array}{lr}
 \hbox{e}^{-(x-\eta_i)^2/2\delta_+^i} & x>\eta_i \\
 \hbox{e}^{-(x-\eta_i)^2/2\delta_-^i} & x\leq\eta_i
 \end{array}
 \right.
\end{equation}
(see e.g. Ref.~\cite{stat}).

\begin{figure}
 \centerline{\epsfig{figure=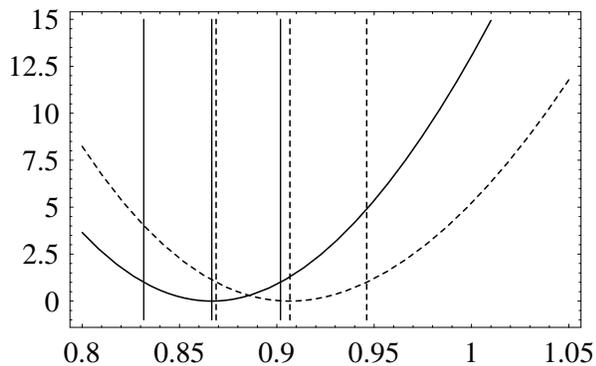,width=8cm}}
 \caption{The function $S$ as a function of $\eta$ for the two data sets (1 in plain and 2 in
 dash).  The vertical bars indicates for each set the position of the
 minimum and the 1$\sigma$ confidence interval defined by $\Delta
 S=1$.}
 \label{fig3}
\end{figure}

\begin{figure}
 \vskip0.2cm
 \centerline{\epsfig{figure=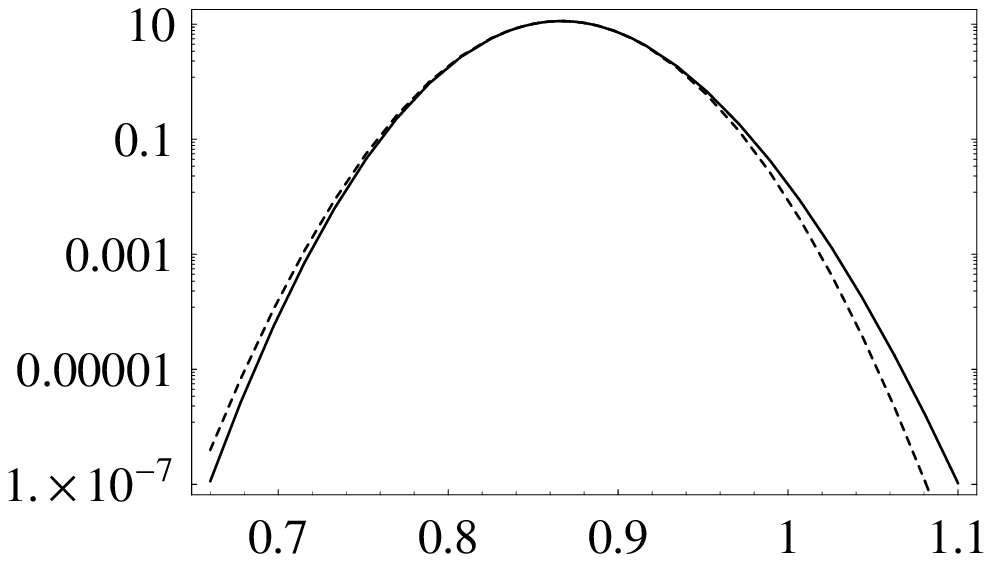,width=4cm}
 \epsfig{figure=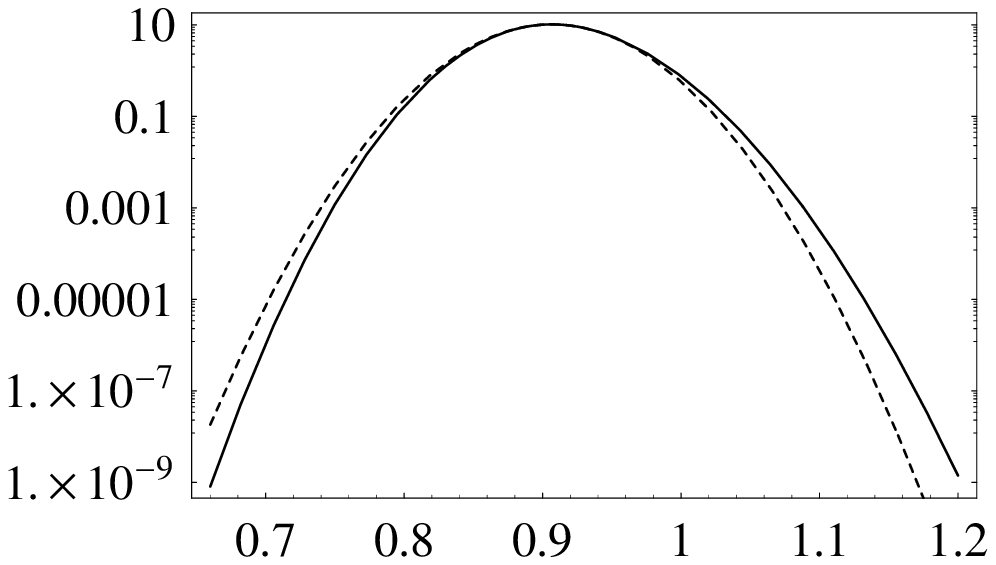,width=4cm}}
 \vskip0.2cm
 \caption{Distribution of $\eta$ obtained from the data of
 Fig.~\ref{fig1} compared with a Gaussian fit (dash) (left=set 1 and
 right= set 2).}
 \label{fig2}
\end{figure}

We perform this analysis using two data sets. The first set,
labelled 1, contains all the clusters, while in the second,
labelled 2, we have removed the point at $z=0.374$ that lies
outside of the other data points. This point corresponds to Abell
370 that clearly shows an apparent bimodal shape in optical and
X-ray data, making its modelling as a single spherical potential a
likely over-simplification for our purpose. The result is
summarized on the plot~\ref{fig3} where we have displayed the
function $S$, its minimum and the 1$\sigma$ confidence level.
Fig.~\ref{fig2} compares the probability distribution function of
$\eta$ to a Gaussian distribution fitted to the data. We obtain
from our analysis that
\begin{equation}
 \eta = 0.87^{+0.04}_{-0.03}
\end{equation}
for the first data set and
\begin{equation}
 \eta = 0.91^{+0.04}_{-0.04}
\end{equation}
for the second.

Additionally we have analyzed, for the second data set (i.e.
without Abell 370), separately the low redshift ($z<0.3$) data and
the high redshift data.  We find for the low redshift set $\eta =
0.89_{-0.05}^{+0.05}$ and for the high redshift set $\eta =
0.95_{-0.07}^{+0.07}$. It is noteworthy that the largest departure
from $\eta=1$ is at small redshifts. These results and the one
obtained for the whole data set suggest that there is no
significant violation of the distance duality relation from
combined X-ray and SZ measurements.

\section{Perspectives and conclusions}\label{sec_4}

Testing for the distance duality relation and/or the reciprocity
relation can give some insight on the puzzling apparent
acceleration of the universe derived from cosmological
observations. To distinguish between various models, one needs
several complementary tests to the reconstruction of the Hubble
parameter as a function of the scale factor (or equivalently of
the equation of state). Examples of such tests are the test of the
Poisson equation on large scales, the test of the constancy of the
fundamental constants and the test of the reciprocity relation.
Figure~\ref{fig0} illustrates how the combination of these tests
can help in identifying  class of models.

We have then shown that observations of galaxy clusters offer a
test of the distance duality relation. In particular, using SZ and
X-ray measurements of the same clusters give an estimate of
$D_A/\eta^2$. An important consequence is that X-ray/SZ combined
analysis does not give a measurement of the angular distances when
the distance duality relation is violated.

Testing the distance duality relation was already proposed in
Ref.~\cite{basset_kunz}. In that work, different sets of data were
used such as type Ia supernovae data to get the luminosity
distance, and the FRIIb radio galaxies, X-ray clusters and compact
radio data to derive angular distances. Interestingly, a general
three parameter form of $\eta(z)$ was proposed in
Ref.~\cite{basset_kunz}, based on general arguments about the
violation of the conservation of the number of photons.  Using
their data, \cite{basset_kunz} found a 2$\sigma$ violation of the
distance duality relation, mainly caused by an excess brightening
of SNIa at redshift larger than 0.5. This analysis also allowed to
put constraints on systematic effects, such as SNIa extinction or
evolution, that may bias apparent magnitudes.

The analysis of the Reese {\em et al.}~\cite{reese} cluster
catalogue has shown that $\eta=1$ is marginally consistent with
the data.  In our study, we have not searched for a fit of a
general expression for $\eta(z)$. Our main concern was first to
use X-ray and SZ combined measurements of galaxy clusters to check
whether the critical value $\eta=1$ was compatible with the data.
Although we found that a value of $\eta$ sightly lower than 1 is
favored, drawing any conclusion on the possible discrepancy
between the distances as predicted in the concordance model and
those determined by our X-ray/SZ combined analysis is premature.
There are indeed several systematic effects that may bias our
derivation of $\eta(z)$, like  over-simplification of cluster
symmetry (substructures, tri-axiallity), or of their temperature
and luminosity radial profiles. For example, the clusters that
deviate most from the $\eta(z)=1$ line on Fig.2 are those that
clearly show bimodal structures from X-ray, SZ and optical images
(Abell 370, Abell 773 and Abell 1689). In contrast, those showing
a single emission region with spherical shape lie very close to
$\eta(z)=1$. It is interesting to note that when the three most
bimodal clusters are removed from the sample, the Reese {\em et
al.}~\cite{reese} remaining clusters lead to
\begin{equation}
 \eta = 0.93^{+0.05}_{-0.04}
\end{equation}
which is compatible to $\eta=1$ at a 2$\sigma$ level. The shape,
temperature distribution etc..., are key points to control in
order to reduce the systematics that limit the accuracy of this
test.

Therefore, we will have to make sure that the marginal
trend\footnote{Note also that this trend is opposite to the one
obtained in Ref.~\cite{basset_kunz} (see their Fig.~1), which
strengthens that there is in fact no systematic trend toward
$\eta<1$.} $\eta(z)<1$ survives further explorations of this
method with new data. This trend is indeed related to the fact
that X-ray/SZ analysis systematically favors a rather low value of
the Hubble constant. Besides, the analysis of the low and high
redshift subsets and the fact that the largest departure from
$\eta=1$ is at low redshift suggest that there is no violation of
the distance duality relation. In particular, parametric forms,
such as the one proposed in Ref.~\cite{basset_kunz}, predict a
cumulative effect with redshift.

More specifically, a larger number of clusters spread over the
whole redshift range and showing simple apparent geometry (i.e. as
compact and spherical as possible) must be selected carefully. It
will improve to lower systematics, to reduce  errors bars on
cluster data and, in turn, to provide much better angular distance
estimates, making the test of the distance duality relation from
X-ray and SZ measurement an efficient method.

\vskip0.5cm {\bf Acknowledgements:} We thank M. Arnaud, F.
Bernardeau, G. Esposito-Far\`ese,  P. Peter, S. Prunet anf F.
Vernizzi for discussions. We also thank B. Bassett for commenting
on its work. This study was motivated by discussions during the
JDEM workshop ({\tt http://www2.iap.fr/pnc/JDEM/}).


\end{document}